\documentclass[preprint,prc,showpacs,preprintnumbers,amsmath,amssymb]{revtex4-2}
\usepackage{graphicx,latexsym,amssymb,amsmath,color,mathrsfs,booktabs,ifpdf}
\usepackage{epsfig,natbib}
\def\bm#1{\mbox{\boldmath $#1$}}
\def\cc{$^{12}$C+$^{12}$C\ }
\def\oo{$^{16}$O+$^{16}$O\ }
\def\AA{nucleus-nucleus\ }
\begin{document}
\title{Comparative folding-model study of low-energy elastic scattering and fusion
 of the $^{12}$C+$^{12}$C and $^{16}$O+$^{16}$O systems}
\author{Le Hoang Chien$^{1,2}$}
\author{Dao T. Khoa$^3$}
\author{Nguyen Hoang Phuc$^{4,2}$}
\author{Doan Thi Loan$^3$}
\author{Nguyen Tri Toan Phuc$^{1,2}$} 
\affiliation{$^1$Department of Nuclear Physics, Faculty of Physics and 
 Engineering Physics, University of Science, 227 Nguyen Van Cu, Cho Quan, Ho Chi Minh City, Vietnam \\
$^2$Vietnam National University Ho Chi Minh City, Linh Xuan, Ho Chi Minh City, Vietnam\\
$^3$Institute for Nuclear Science and Technology, VINATOM, 179 Hoang Quoc Viet, Nghia Do, Hanoi 122772, Vietnam\\
$^{4}$Department of Applied Physics, Faculty of Applied Science, Ho Chi Minh City University of Technology 
 (HCMUT), 268 Ly Thuong Kiet, Dien Hong, Ho Chi Minh City, Vietnam \\ }
\date{\today}
\begin{abstract}
A density dependent nucleon-nucleon interaction (CDM3YR) has been parametrized based on the original 
M3Y-Reid interaction, to properly reproduce the saturation properties of symmetric nuclear matter (NM) 
in the nonrelativistic Hartree-Fock calculation, with the energy of NM in a good agreement with the 
microscopic \emph{ab-initio} results over densities up to three times the saturation density. The real 
optical potential (OP) of symmetric \cc and \oo systems is then calculated within the double-folding 
model (DFM), using the realistic densities of $^{12}$C and $^{16}$O nuclei and CDM3YR interaction, 
for the optical model analysis of elastic scattering at low energies and determination of the astrophysical 
$S$ factor of  \cc and \oo fusion in the barrier penetration model. The DFM calculation of the real OP 
for these two symmetric systems was also done using the original (density independent) M3Y-Reid 
interaction, and that added by a repulsive core suggested by Esbensen {\it et al.} \cite{Esb08,Esb11}, 
to explore the impact of medium effects that are effectively encoded in the density dependence 
of the CDM3YR interaction.   
\end{abstract}
\maketitle

\section{Introduction}\label{intro} 
The investigation of \cc and \oo fusion is of high importance for astrophysical studies of structure and 
evolution of massive stars. The stellar nucleosynthesis starts from the burning of four hydrogen atoms 
to form helium, to the helium burning to form carbon and oxygen, and then proceeds, depending 
on the star mass, to the carbon- and oxygen burning phase to produce heavier nuclei 
\cite{Fow84,Rol88,Ili15}. For a massive star with $M \gtrsim 8M_\odot$ where gravitational force 
overcomes the pressure of degenerate electrons and gravitational potential energy converts partially 
into heat to avoid stellar collapse, so that temperature becomes high enough to ignite \cc fusion 
to form $^{20}$Ne, $^{23}$Na, and $^{23}$Mg through reactions  
$^{12}$C($^{12}$C,$\alpha$)$^{20}$Ne ($Q =$ 4.617 MeV),
$^{12}$C($^{12}$C,$p)^{23}$Na ($Q =$ 2.241 MeV), 
$^{12}$C($^{12}$C,$n)^{23}$Mg ($Q = -$2.6 MeV). 
The stellar condition for \cc fusion requires temperature of about $(6\sim 8)\times 10^8$ K and 
mass density around $10^5$ g~cm$^{-3}$. With a higher Coulomb barrier, \oo fusion (the main 
reaction during the oxygen burning phase) requires higher temperature of about 
$(1.5\sim 2.7)\times 10^9$ K and mass density around $(2.6\sim 6.7)\times 10^9$ g~cm$^{-3}$. 
The \oo fusion forms the compound nucleus $^{32}$S that decays through various particle-emission 
channels  
$^{16}$O($^{16}$O,$p)^{31}$P ($Q = 7.678$ MeV),
$^{16}$O($^{16}$O,$2p)^{30}$Si ($Q = 0.381$ MeV),
$^{16}$O($^{16}$O,$\alpha)^{28}$Si ($Q =9.594$ MeV),
$^{16}$O($^{16}$O,$d)^{30}$P ($Q =-2.409$ MeV), 
$^{16}$O($^{16}$O,$n)^{31}$S ($Q = 1.499$ MeV) \cite{Ili15}. 

Because of the vital role of  \cc and \oo fusions in stellar nucleosynthesis, numerous experimental 
and theoretical studies of these fusion reactions were carried out at low energies \cite{Mat69,Spi74,
Hul80,Wu84,Tho86,Kur87,Dua15,Gas07,Esb08,Koc10,Pat69,Maz73,Hig77,Das82,Agu06,Bar06,
Gas05,Spi07,Buc15,Den10,Jia13,Ass13,Azi15,Jia18,Tum18,Chi25,Esb11,Chi18,God19,Sim13,Des26}. 
The main challenge for the experimental studies is that stellar \cc and \oo fusions occur at very low 
energies of about 1.5 MeV and 6 MeV, respectively, which are well below the Coulomb barriers. This 
makes measurements very difficult to obtain the fusion cross section with high accuracy at energies 
close to stellar conditions \cite{Jia18,Tum18}. In fact, there is a considerable discrepancy between 
the \cc fusion cross sections obtained from the direct and indirect measurements \cite{Tum18,Chi25}. 
Similarly, \oo fusion cross sections at energies below the lowest measured energy of 6.8 MeV 
obtained by the $\gamma$ techniques are by a factor of 3 larger than that given by the particle 
techniques \cite{Spi74}. Therefore, a reliable theoretical prediction of  \cc and \oo fusion cross 
sections at astrophysical energies should be of interest for experimental studies. To this goal, 
various potential model studies were done either in the one-channel barrier penetration model 
(BPM) or the coupled channel formalism \cite{Bal98}, and the key input is the model of 
nuclear potential chosen  for the fusion calculation 
\cite{Esb08,Esb11,Koc10,Gas05,Den10,Ass13,Azi15,Chi18,God19,Sim13}.

Such potential models can be roughly divided into two groups. The first group refers to the \AA 
potential parametrized phenomenologically in some functional form like, e.g., the Woods-Saxon 
potential. The second group refers to the \AA potential evaluated microscopically starting 
from nucleon degrees of freedom. A widely used approach is the double folding model (DFM) 
calculation of the \AA potential \cite{Bra97,Kho00,Gas05,Gas07,Chi18}, using an effective 
nucleon-nucleon (NN) interaction and appropriate nuclear densities. We note, in particular, 
a hybrid folding model approach suggested by Esbensen {\it et al.} \cite{Mis06,Mis07,Esb08,Esb11}, 
where the folded \AA potential given by the original (density-independent) M3Y-Reid interaction 
\cite{Ber77} is supplemented by a phenomenological repulsive term, referred to hereafter as the M3Y+Rep 
potential. The nuclear energy density functional approach was also used to determine the \AA interaction 
potential of \cc and \oo systems \cite{God19,Sim13}. In addition to the potential models, a more 
comprehensive microscopic study of \cc fusion was carried out recently based on the resonating group 
method (RGM), with the effective NN interaction for the RGM calculation as the only input, 
and $\alpha$-emission channel $^{12}$C($^{12}$C,$\alpha$)$^{20}$Ne treated explicitly \cite{Des26}. 

Guided by the consistent mean-field description of light heavy-ion (HI) optical potential (OP) at medium
and low energies \cite{Bra97,Kho07r}, it was shown \cite{Chi18} that the \cc interaction potential 
given by the DFM belongs to the deep family of the real OP that gives both good optical model 
(OM) description of elastic \cc scattering at low energies and BPM description of \cc fusion. 
The potential-model study of  molecule states of $^{24}$Mg or $^{32}$S nuclei \cite{Buc90,Ohk02} 
also suggested that the corresponding \cc or \oo potential should be deep enough to provide a 
proper explanation of the underlying band of  di-cluster resonances \cite{Kon98}. With energies 
increasing to around the Fermi energy, the real OP given by the DFM was shown to account 
properly for the nuclear rainbow pattern observed in elastic light HI scattering 
\cite{Bra97,Kho05,Kho07r,Kho16}. The evolution of the nuclear rainbow pattern in 
symmetric \cc and \oo systems is strongly affected by the boson symmetry of these systems, and 
can be described only by the (mean-field type) real folded OP \cite{Phu24}. It is, therefore, 
of interest to explore in more details the validity of the real folded OP in the description of  
elastic \cc and \oo scattering at low energies and fusion reaction.

Because the low-energy elastic scattering data at large angles are quite sensitive to the real OP at 
overlapping distances of two colliding nuclei  \cite{Chi18,Kho07r,Kho16,Phu24}, the medium dependence 
of the effective NN interaction must be taken into consideration. In general, the medium effects 
are usually taken into account in the DFM calculation by using a \emph{density dependent} NN 
interaction. It is naturally expected that the mean-field Hartree-Fock (HF) calculation of NM using 
a realistic density dependent NN interaction reproduces properly the saturation properties of cold 
nuclear matter (NM). In the present comparative folding model study, we have parametrized a new 
density dependent NN interaction based on the original M3Y-Reid interaction \cite{Ber77}, dubbed 
hereafter as the CDM3YR interaction, with a rearrangement term determined 
from the extended HF calculation of nucleon OP in NM \cite{Kho16}.  
        
\section{Density dependent CDM3YR interaction and HF calculation of NM}
\label{sec1} 
We briefly recall the (nonrelativistic) Hartree-Fock description of  homogeneous and symmetric NM 
at given nucleon number density $\rho$. In this case, the energy density of NM inside a macroscopic 
volume $\Omega$ is obtained with the direct ($v^{\rm D}_{00}$) and exchange ($v^{\rm EX}_{00}$) 
parts of the (spin- and isospin independent) in-medium NN interaction as 
\begin{equation}
\mathcal{E}=\frac{1}{ \Omega}\sum_{k \sigma \tau}\frac{\hbar^2k^2}{2m}+
\frac{1}{ 2 \Omega}\sum_{k \sigma \tau}\sum_{k'\sigma '\tau '}
[\langle{\bm k}\sigma \tau, {\bm k}' \sigma' \tau' |v^{\rm D}_{00}|{\bm k}\sigma\tau,
{\bm k}' \sigma' \tau' \rangle +\langle{\bm k}\sigma \tau, {\bm k}'\sigma' \tau' |v^{\rm EX}_{00}|
{\bm k}'\sigma \tau, {\bm k}\sigma' \tau' \rangle], \label{eq1} 
\end{equation} 
where $m$ is the nucleon mass, and $|{\bm k}\sigma \tau\rangle=\displaystyle\frac{\exp{({i{\bm k}{\bm r})}}}
{\sqrt{\Omega}}\chi_\sigma\chi_\tau$ are plane waves. The earlier
density dependent interactions based on the original M3Y interaction \cite{Ber77} were employed 
successfully in both the HF studies of NM and the folding model calculations of the \AA OP 
\cite{Kho93,Kho97}. For the \emph{spin-saturated} symmetric NM, the explicit spin- and isospin 
independent part of the density dependent M3Y interaction \cite{Kho97} is    
\begin{equation}
 v^{\rm D(EX)}_{00}(\rho, s)=F_{00}(\rho)v^{\rm D(EX)}_{00}(s) 
=C[1+\alpha\exp(-\beta \rho) + \gamma \rho]~v^{\rm D(EX)}_{00}(s), \  s=|{\bm r}-{\bm r}'|. 
\label{eq2}
\end{equation}
Here $s$ is the internucleon separation, and the radial part of the original M3Y-Reid interaction is 
determined in term of three Yukawas \cite{Ber77}  
\begin{align}
 v^{\rm D}_{00}(s) = 7999.0 \frac{{\rm exp}(-4s)}{4s}
  - 2134.25 \frac{{\rm exp}(-2.5s)}{2.5s},\hskip 3 cm \nonumber\\
 v^{\rm EX}_{00}(s) = 4631.38 \frac{{\rm exp}(-4s)}{4s} 
 - 1787.13 \frac{{\rm exp}(-2.5s)}{2.5s}
 - 7.8474 \frac{{\rm exp}(-0.7072s)}{0.7072s}. \label{eq3}
\end{align}
The nuclear energy density (\ref{eq1}) of symmetric NM at the nucleon number density $\rho$ is then 
obtained \cite{Kho93,Kho94,Kho97} as  
\begin{equation}
\mathcal{E}(\rho)=\frac{3\hbar^2k^2_F\rho}{10m}+\frac{\rho^2F_{00}(\rho)}{2}
\left\{J^{\rm D}_{00}+ \int\left[\widehat{j}_1(k_F s)\right]^2 v^{\rm EX}_{00}(s) d^3s\right\},
 \label{eq4} 
\end{equation} 
where $k_F=(1.5\pi^2\rho)^{1/3}$, $J^{\rm D}_{00}=\displaystyle\int v^{\rm D}_{00}(s)d^3s$ is 
the volume integral of the direct radial interaction (\ref{eq3}), and $\widehat{j}_1(x)=3j_1(x)/x$. 
The empirical saturation properties of symmetric NM imply the minimum of energy per particle 
$\displaystyle\frac{E}{A}(\rho)\equiv \displaystyle\frac{\mathcal{E}(\rho)}{\rho}\approx -16$ MeV at
the saturation density $\rho_0 \approx 0.17$ fm$^{-3}$. The equation of state (EOS) of symmetric 
NM is usually characterized by the pressure of NM 
\begin{equation}
  P(\rho)=\rho^2\frac{\partial}{\partial\rho}\frac{E}{A}(\rho), \label{eq5} 
\end{equation} 
and nuclear incompressibility $K$ at the saturation density 
\begin{equation}
  K=9\frac{\partial P(\rho)}{\partial\rho}\Big\arrowvert_{\displaystyle{\rho\to \rho_0}}=
	9\rho^2\frac{\partial^2}{\partial\rho^2}\frac{E}{A}(\rho)\Big\arrowvert_{\displaystyle{\rho\to \rho_0}}.
	\label{eq6} 
\end{equation}   
We have adjusted in the parameters of the density functional $F_{00}(\rho)$ in Eq.~(\ref{eq2}), used with 
the radial M3Y-Reid interaction (\ref{eq3}) to properly reproduce the saturation point of symmetric NM 
and the nuclear incompressibility $K\approx 223$ MeV (see explicit parameters of the CDM3YR interaction 
in Table~\ref{t1}). 

\begin{figure}[bht]\vspace{-2cm} 
\hspace{1.0cm}\vspace{-3.0cm}
\includegraphics[angle=0,scale=0.65]{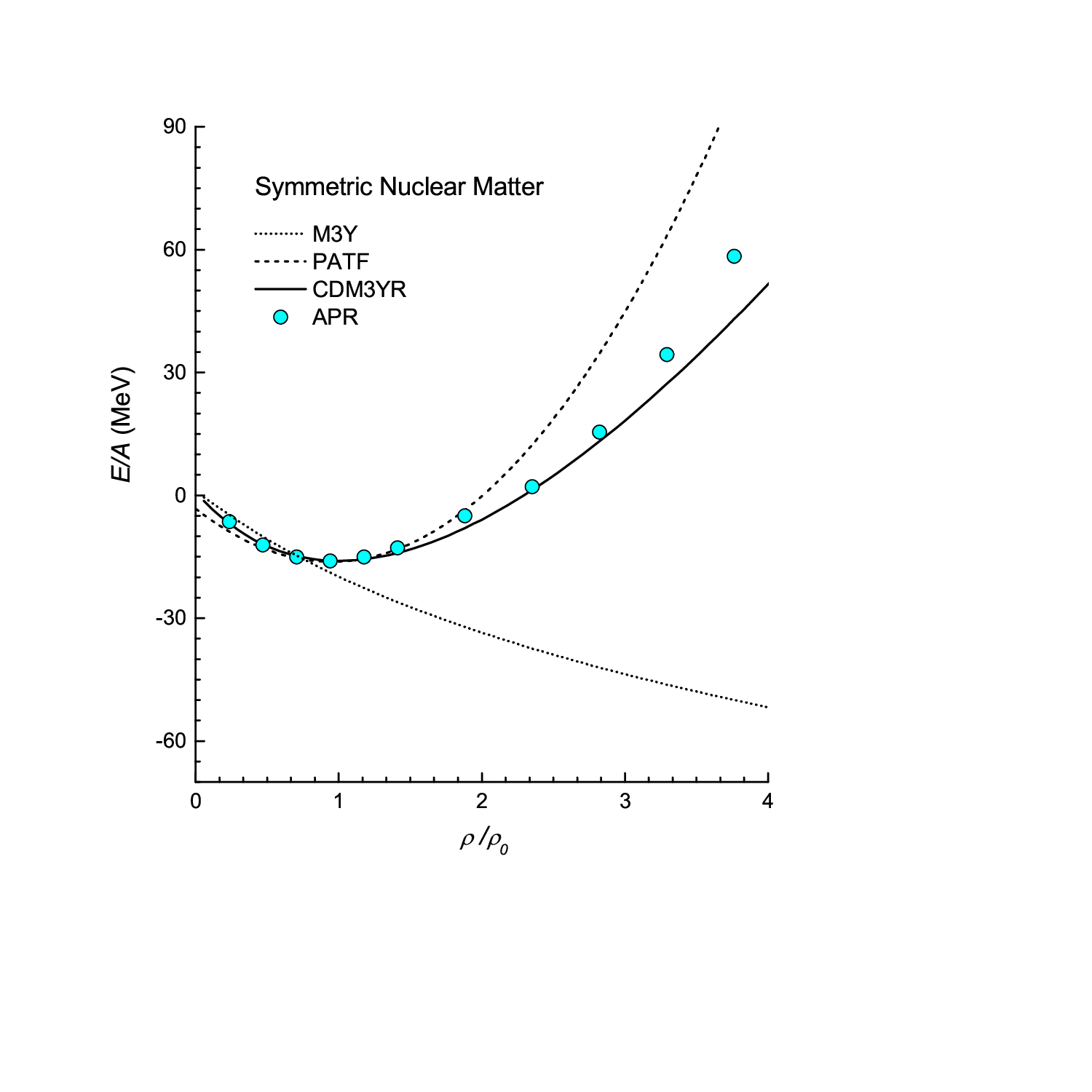}\vspace{-0.5cm}
\caption{Energy per particle $E/A$ of symmetric NM given by three approaches: the PATF model 
(dashed curve), HF calculation using the original (density independent) M3Y-Reid interaction 
(dotted curve) and the new density dependent CDM3YR interaction (solid curve). The solid circles 
are the results of the ab-initio variational calculation APR \cite{Akm98}.} \label{fig1}
\end{figure}

The HF results given by the CDM3YR density dependent interaction are plotted in Fig.~\ref{fig1} as solid 
curve, in comparison with results of the same HF calculation but using the original (density independent) 
M3Y-Reid interaction (\ref{eq3}) shown as dotted curve. Lacking a realistic density dependence, the original 
M3Y-Reid interaction fails to reproduce the saturation of  NM, and results on collapsing NM as discussed 
earlier in Ref.~\cite{Kho93}.  These HF results are also compared with those of the \emph{ab-initio} 
variational calculation using A18 version of the Argonne NN potential added by a three-body force 
by Akmal, Pandharipande and Ravenhall (APR) \cite{Akm98}, and energy of symmetric NM given by the 
Thomas-Fermi model (PATF curve in Fig.~\ref{fig1}) \cite{Mis07,Mye98} that is parametrized in a 
parabolic form as
\begin{equation}
 \frac{E}{A}(\rho) = \frac{E}{A}(\rho_0) + \frac{K}{18\rho_0^2} (\rho-\rho_0)^2, \label{eq7}
\end{equation}
where $\displaystyle\frac{E}{A}(\rho_0)\approx -16$ MeV and $K\approx 234$ MeV \cite{Esb11}. 
The parabolic EOS (\ref{eq7}) was used to validate the phenomenological repulsive term used in the 
folding model calculation with the M3Y+Rep interaction \cite{Mis06,Mis07,Esb08,Esb11}. 

\begin{table}
\caption{Parameters of the density functionals $F_{00}(\rho)$ and $\Delta F_{00}(\rho)$ 
of the CDM3YR interaction. $K$ is the incompressibility of symmetric NM at the saturation 
density  (\ref{eq6}).} \vskip 1cm
\begin{tabular}{|c|c|c|c|c|c|c|} \hline
                & $C$   & $\alpha$ & $\beta$ & $\gamma$ & $K$  \\
  &  &   & (fm$^3$) & (fm$^3$) & (MeV) \\ \hline
 $F_{00}(\rho)$        & 0.2535 & 3.8801   & 1.0101  & $-$4.0     & 223  \\
 $\Delta F_{00}(\rho)$ & 1.939 & 1.0      & 0.587   &      & \\ \hline
\end{tabular}\label{t1}
\end{table} 
Except the results given by the density independent M3Y-Reid interaction, those of the other three 
approaches shown in Fig.~\ref{fig1} describe the saturation point equally well but give rather  
different slopes of the NM energy at higher densities. In particular, the parabolic form adopted for PATF 
(\ref{eq7}) predicts a much stiffer $E/A$ curve compared to two remaining methods. The HF results 
obtained with the newly parametrized CDM3YR interaction agree closely with the \emph{ab-initio} 
APR results over NM densities up to  $3\rho_0$, and the use of the CDM3YR interaction is thus 
well validated for the low-energy domain of the nuclear EOS. 

To account for the rearrangement term (RT) of the single-nucleon potential in NM, implied naturally 
by Hugenholtz-van Hove theorem, the density dependence (\ref{eq2}) of the CDM3YR interaction was 
extended and modified in the same mean-field manner as discussed in Ref.~\cite{Kho16} to account 
for the RT in the DFM calculation of \AA OP     
\begin{align}
 v^{\rm D(EX)}_{00}(\rho,s)=F_{00}(\rho)v^{\rm D(EX)}_{00}(s)\to 
 [ F_{00}(\rho)+\Delta F_{00} (\rho) ]v^{\rm D(EX)}_{00} (s), \hskip 1.5cm \label{eq8} \\
\Delta F_{00}(\rho) = C [ \alpha~\exp(-\beta \rho) -1 ], \ 
 \mbox{with parameters given in Table ~\ref{t1}}. \nonumber
\end{align}

\section{Double-folding model of the real nucleus-nucleus OP}
\label{sec2} 
\subsection*{DFM calculation with the density dependent CDM3YR interaction} 
For symmetric \cc and \oo systems under study, the real \AA OP is evaluated as the HF-type 
potential \cite{Kho97,Kho07r} using the density dependent CDM3YR interaction 
(\ref{eq8})  with the RT included 
\begin{equation}
  V_{\rm N}=V_{\rm D}+V_{\rm EX}=\sum_{i\in A_1,j\in A_2}[\langle ij|v^{\rm D}_{00}|ij\rangle 
	+ \langle ij |v^{\rm EX}_{00}|ji\rangle]. \label{eq9}
\end{equation}
Here $|i\rangle$ and $|j\rangle$ are the single-particle wave functions of  projectile ($A_1$) 
and target ($A_2$) nucleons, respectively. The \emph{local} direct part of the internuclear 
potential (\ref{eq9}) is determined with the ground-state (g.s.) densities of two colliding nuclei as 
\begin{equation}
 V_{\rm D}(R)=\int\rho_{A_1}({\bm r}_1)\rho_{A_2}({\bm r}_2)
 v^{\rm D}_{00}(\rho,s)d^3r_1 d^3r_2, \  {\bm s}={\bm r}_1-{\bm r}_2+{\bm R}.
\label{eq10}
\end{equation}
The antisymmetrization takes into account explicitly the (pair-wise) knock-on exchange between 
projectile- and target nucleons. As a result, the exchange term of $V_{\rm N}$ becomes 
\emph{nonlocal} in coordinate space \cite{Kho07r}. An accurate local approximation is usually made 
by treating the relative motion locally as a plane wave \cite{Kho07r}, so that the exchange term 
of $V_{\rm N}$ is obtained in the following local form 
\begin{eqnarray}
V_{\rm EX}(R)&=&\int\rho_{A_1}({\bm r}_1,{\bm r}_1 +{\bm s})
 \rho_{A_2}({\bm r}_2,{\bm r}_2 -{\bm s}) \nonumber \\ 
 &\times& v^{\rm EX}_{00}(\rho,s)\exp\left(\frac{i{\bm K}(E,R).{\bm s}}{M}
\right)d^3r_1d^3r_2, \label{eq11}
\end{eqnarray}
where $\rho_{A_{1(2)}}({\bm r},{\bm r}')$ are the nonlocal density matrices of  two colliding
nuclei, $M=A_1A_2/(A_1+A_2)$ is the recoil factor, $A_1$ and $A_2$ are the projectile- 
and target mass numbers, respectively. The local relative-motion momentum $K(E,R)$ is determined 
self-consistently from the real \AA OP (\ref{eq9}) as 
\begin{equation}
 K^2(E,R)={{2\mu}\over{\hbar}^2}[E-V_{\rm D}(R)-V_{\rm EX}(R)-V_{\rm C}(R)], \label{eq12}
\end{equation}
where $\mu$ is the reduced mass, $E$ is the center-of-mass (c.m.) energy, and Coulomb potential 
$V_{\rm C}$ is obtained by folding Coulomb force with the charge densities of two colliding nuclei, 
chosen to have the charge radii $R_C$ = 3.17 fm and 3.54 fm for $^{12}$C and $^{16}$O, 
respectively. 

Because the strength and slope of the folded nuclear potential (\ref{eq9})-(\ref{eq12}) at sub-surface radii 
are quite sensitive to the \AA overlap density, a proper treatment of the overlap density  $\rho$ appearing 
in (\ref{eq8}) is important in the present DFM calculation. The frozen density approximation (FDA), 
which adopts the sum of two frozen densities ($\rho=\rho_1+\rho_2$) for the overlap density, is widely 
used in the DFM calculations for energies $E>10$ MeV/nucleon. At lower energies, the compression 
of the nuclear overlap is much weaker and an adiabatic density approximation (ADA) is more appropriate 
\cite{Chi18,Oht87,Ich15}. Such an ADA is adopted in the present DFM calculation with the CDM3YR 
interaction (see more details in Ref.~\cite{Chi18}), where the overlap density of the \cc or \oo system 
gradually changes with decreasing internuclear distance, from that given by the FDA to the central  
density of $^{24}$Mg or $^{32}$S compound nucleus, respectively.  
\begin{figure}[bht]
\vspace*{-1.5cm}\hspace*{1.5cm}
\includegraphics[angle=0,scale=0.80]{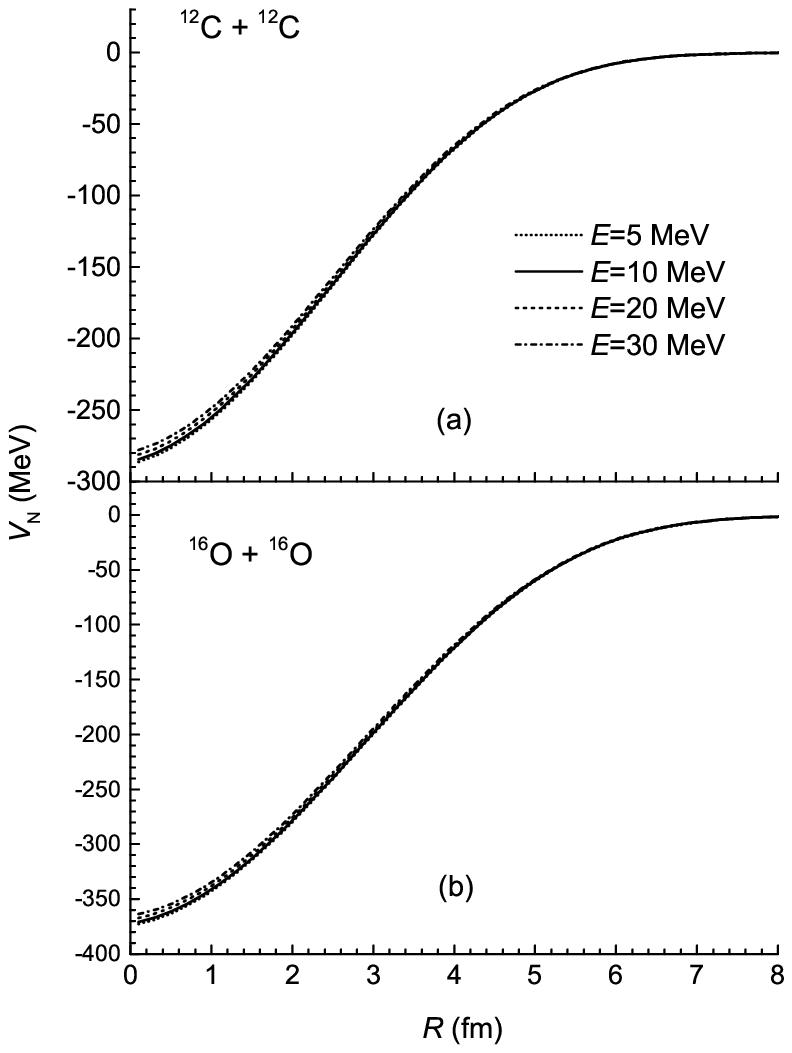}\vspace*{-1.5cm}
\caption{The folded nuclear potential (\ref{eq9}) obtained for $^{12}$C+$^{12}$C (a) and 
$^{16}$O+$^{16}$O (b) systems at c.m. energies $E=5-30$ MeV, using the density dependent 
CDM3YR interaction (\ref{eq8}).} \label{fig2}
\end{figure}

The DFM calculation of the real \AA OP (\ref{eq9}) for  symmetric \cc and \oo systems at energies 
of 5 to 30 MeV was done using realistic g.s densities of $^{12}$C and $^{16}$O nuclei given 
by the no-core shell model \cite{Gen18} and cluster variational Monte-Carlo calculations \cite{Lon96},
respectively, and the results are shown in Fig.~\ref{fig2}. One can see that the real folded potential 
has the depth of about -280 MeV and -370 MeV for \cc and \oo systems, respectively. Although the 
folded potential (\ref{eq9}) is energy dependent because of the localized exchange term 
(\ref{eq11})-(\ref{eq12}), the potential depth varies weakly in the energy range of 5 to 30 MeV. 
We have chosen, therefore, the folded potential obtained at $E=10$ MeV for each of these two systems 
in all OM and BPM calculations discussed in the present work. 

\subsection*{DFM calculation with the M3Y+Rep interaction}
Recently, a hybrid folding model based on the M3Y+Rep interaction \cite{Esb08,Esb11} was used 
to study fusion of  symmetric \cc and \oo systems as well as heavier ion systems. In this approach, 
a strong \emph{repulsive} core is introduced to the nuclear potential given by the DFM calculation
using the density independent M3Y-Reid interaction \eqref{eq3}, supposedly to \emph{repel} 
the overlap of the frozen wave functions of two colliding nuclei at small distances \cite{Mis06,Mis07}. 
In such a hybrid approach, 
\begin{equation}
 V_{\rm N}(R) = V_{\rm M3Y}(R) + V_{\rm Rep} (R), \label{eq13}
\end{equation}
where $V_{\rm M3Y}(R)$ is the folded \AA potential given by the direct M3Y-Reid 
interaction (\ref{eq3}), using a zero-range pseudo-potential for the knock-on exchange   
\begin{equation}
 V_{\rm M3Y}(R) = \int \rho_{A_1} (\bm{r}_1) \rho_{A_2} (\bm{r}_2)
 \left[ v^{\rm D}_{00}(s)+ \hat{J}_{00} \delta (s)\right] d^3 r_1 d^3 r_2, \label{eq14}
\end{equation}
where $\hat{J}_{00}=276$ MeV~fm$^3$ \cite{Sat79}. The repulsive term in Eq.~(\ref{eq13}) 
is obtained by double folding two nuclear densities with a zero-range repulsive force
\begin{equation}
 V_{\rm Rep}(R) = \int \rho_{A_1}(\bm{r}_1) \rho_{A_2} (\bm{r}_2) v_{\rm Rep}
 \delta (s) d^3 r_1 d^3 r_2. \label{eq15}
\end{equation}
While the realistic two-parameter Fermi density was used to evaluate $V_{\rm M3Y}(R)$
 \cite{Esb08,Esb11,Mis06,Mis07}, the diffuseness of the Fermi density was further calibrated 
with the repulsive strength $v_{\rm Rep}$ in the calculation of the repulsive potential (\ref{eq15}), 
so that the nuclear folded potential (\ref{eq13}) at zero distance approximately equals the 
difference of the NM energy per particle as density increases from $\rho_0$ to $2\rho_0$. 
From the parabolic EOS (\ref{eq7}), the repulsive core of the nuclear potential is obtained 
from the nuclear incompressibility $K$ scaled by the projectile mass number as  
\begin{equation}
 V_{\rm N} (R=0) \approx 2A_1 \left[ \frac{E}{A}(2\rho_0)-\frac{E}{A}(\rho_0) \right] = 
 \frac{A_1}{9}K. \label{eq16}
\end{equation}
\begin{figure}[bht]
\vspace{-1.0cm}
\hspace*{2.5cm}
\includegraphics[angle=0,scale=0.70]{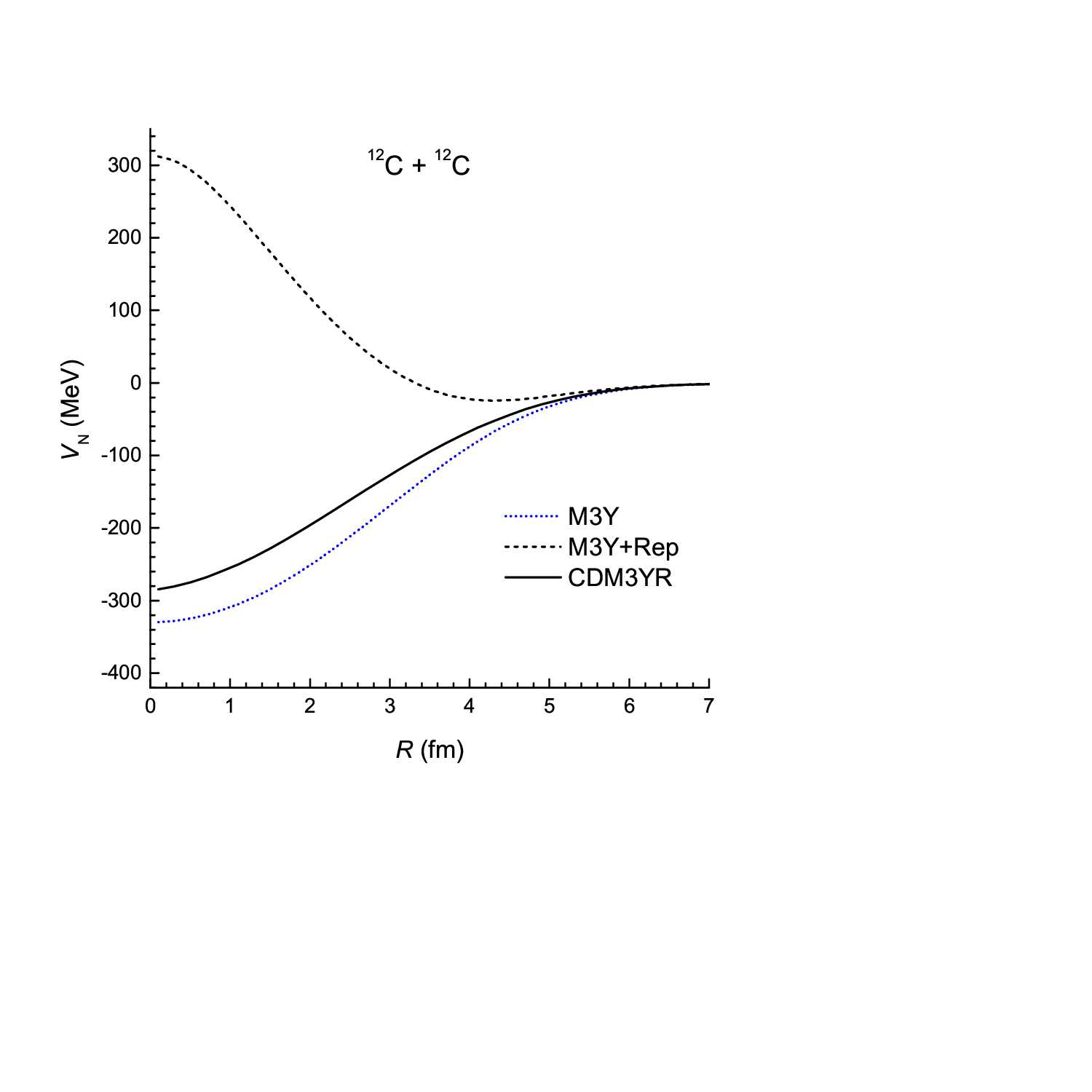} \vspace{-5.5cm}
\caption{The folded nuclear potential of  \cc system  at the c.m. energy $E=10$ MeV 
given by the M3Y (dotted curve), M3Y+Rep (dashed curve) and CDM3YR (solid curve) 
interactions.} \label{fig3}
\end{figure}

The parameters taken from the  \cc fusion study by Esbensen {\it et al.} \cite{Esb11} were used in 
the present calculation of the M3Y+Rep potential (\ref{eq14}) for \cc system. Namely, the  
Fermi distribution of the diffuseness $a$ = 0.44 fm and radius $r$ = 2.155 fm was adopted
for the calculation of  $V_{\rm M3Y}(R)$ potential. A smaller diffuseness $a = 0.31$ fm of the Fermi 
density and $v_{\rm Rep}=398.64$ MeV~fm$^3$ were further used to evaluate the repulsive 
term $V_{\rm Rep}(R)$ in Eq.~(\ref{eq15}). These values were obtained from Eq.~ (\ref{eq16})
with the nuclear incompressibility $K= 234$ MeV. The Fermi density parameters $r = 2.5$ fm and 
$a=0.52$ fm adopted in Ref.~\cite{Esb08} were used for \oo system. By constraining to the same $K$ 
value, a smaller diffuseness $a=0.41$ fm and $v_{\rm Rep}= 494.75$ MeV~fm$^3$ were obtained.

As illustration, the nuclear potentials of \cc system at $E=10$ MeV given by the three folding-model 
are compared in Fig.~\ref{fig3}. While the M3Y potential (dotted curve) has a depth of about -390 MeV, 
the CDM3YR potential (solid curve) becomes significantly less attractive at small radii. In contrast to the 
attractive M3Y and CDM3YR folded potentials, the M3Y+Rep potential (dashed curve) becomes 
\emph{repulsive} at $R\lesssim 3$ fm, reaching a strong repulsive core of above +300 MeV in 
the center. Sometimes, such a huge repulsive core was discussed as the Pauli repulsion which seemed 
plausible but was not microscopically validated. We note that the total energy density of NM 
contains both the kinetic- and potential energy densities, i.e., the first and second terms in Eq.~(\ref{eq4}), 
respectively. In a local density approximation, only the potential energy density in Eq.~(\ref{eq4}) should 
be used in modeling the \AA potential, and to constrain $V_{\rm Rep}$ term of the M3Y+Rep folded 
potential by the relation  (\ref{eq16}) is highly questionable. It should also be noted that the kinetic 
energy of the \AA motion is taken explicitly into account in both the OM and BPM calculations discussed 
in the present work, separately from the \AA interaction potential $V_{\rm N}$. In this connection, 
we mention the earlier DFM calculation of the real OP for symmetric \cc and \oo systems using the density 
dependent M3Y-Reid interaction that shows a grossly \emph{attractive} contribution of the Pauli exchange 
term at small distances (see, e.g., Fig.~5 in Ref.~\cite{Kho94}).    

\section{Elastic \cc and \oo scattering at low energies} 
To probe three versions of the folding model discussed above, we perform the OM analysis 
of the elastic \cc and \oo scattering data measured at energies below 6 MeV/nucleon \cite{Rei73}, 
which are still sensitive to the nuclear interaction between the two nuclei. A hybrid complex OP 
is adopted, with the real nuclear and Coulomb potentials given by the double-folding calculation 
and the imaginary absorptive potential parametrized in Woods-Saxon (WS) form  
\begin{align}
 U(R)=V_{\rm N}(R) + i W_V(R)+ i W_D(R)+ V_{\rm C}(R), \label{eq17} \\
 W_V(R)=-\frac{W_V }{1+\exp[(R-R_V)/a_V]}, \label{eq18} \\ 
 W_D(R)=-4 a_D \frac{d}{dR}\left\{\frac{W_D }{1+\exp[(R-R_D)/a_D]} \right\}.
 \label{eq19} 
\end{align}
The real nuclear potential $V_{\rm N}(R)$ given by the three DFM calculations was used 
without any renormalization of its strength. The imaginary OP consists of the WS volume  (\ref{eq18}) 
and surface (\ref{eq19}) terms, with the latter used for elastic \oo scattering only. 
All OM calculations were done using the code ECIS97 written by Raynal \cite{Ray72}.
\begin{figure}[bht]\vspace{0cm}
\hspace*{2.5cm}
\includegraphics[angle=0,scale=0.7]{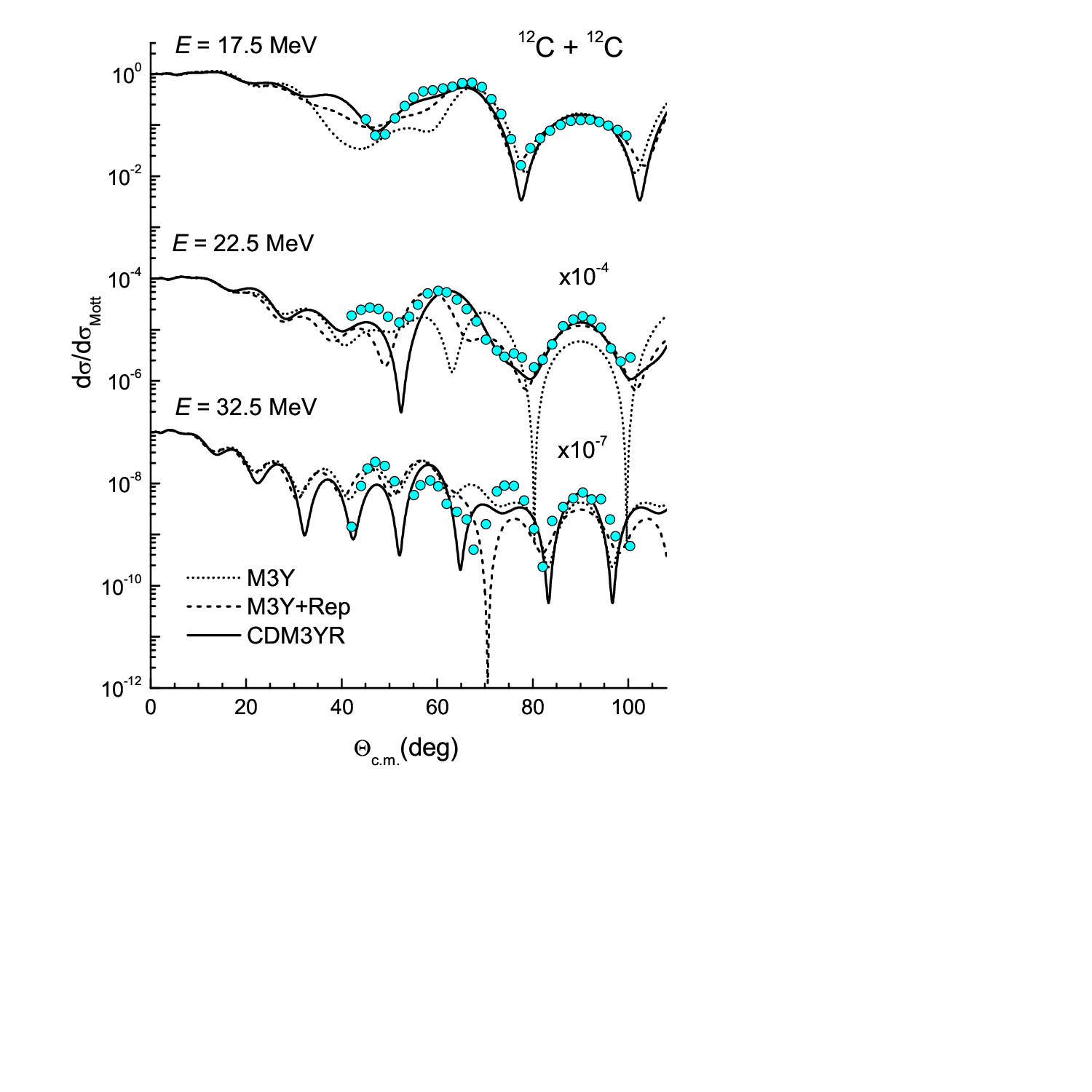}\vspace{-5cm}
\caption{OM description of  elastic $^{12}$C+$^{12}$C scattering at $E = 17.5, 22.5$, and 32.5 MeV 
given by three versions of the real folded OP shown Fig.~\ref{fig3}. The WS parameters (\ref{eq18})
of the imaginary OP were adjusted in each case by the best OM fit to the measured elastic 
data \cite{Rei73}.} \label{fig4}
\end{figure}

\begin{table}[bht]
\caption{\label{tOP1} WS parameters (\ref{eq18}) of the imaginary OP obtained with three 
versions of the folded potential $V_{\rm N}$ from the best OM fit to elastic \cc scattering data 
at $E=17.5, 22.5$, and 32.5 MeV. $J_R$ and $J_W$ are the volume integrals (per interacting 
nucleon pair) of the real and imaginary OP, respectively. $\sigma_R$ is the total reaction cross 
section given by the OM calculation. $\chi^2$ values are per datum, and obtained with uniform 
10\% errors.}\vspace{0.5cm}
  \begin{tabular}{|c|c|c|c|c|c|c|c|c|c|c|c|c|} \hline
 $E $&Potential& $-J_R$    & $W_V$ & $R_V$ & $a_V$ & $-J_W$     & $\sigma_R$ & $\chi^2$\\
(MeV)&         &  (MeV~fm$^3$) & (MeV) & (fm)  & (fm)  & (MeV~fm$^3$) & (mb) & \\ \hline
17.5 &CDM3YR   &    375.5      & 3.172 &6.608  &0.201  &   26.87    &    987.5 & 10.8 \\ \hline
     &M3Y      & 	475.4	   & 2.563 &6.471  &0.203  &   20.40    &    909.9  & 19.2 \\ \hline 
     &M3Y+Rep  & 	23.10	   & 4.138 &6.594  &0.203  &   34.84    &    998.7  & 10.4 \\ \hline
22.5 &CDM3YR   & 	375.5	   & 3.563 &6.402  &0.216  &  27.50     &    1066.0  & 19.1 \\ \hline
     &M3Y      & 	475.4	   & 6.927 &6.341  &0.308  &  52.56     &    1136.3   & 125.1 \\ \hline
     &M3Y+Rep  & 	23.10	   & 4.211 &6.528  &0.394  &  35.32     &    1168.9   & 22.7 \\ \hline
32.5 &CDM3YR   & 	375.5	   & 8.549  &6.100  &0.300  &   57.79    &    1188.3   & 153.5  \\ \hline
     &M3Y      & 	475.4	   & 10.927&6.336  &0.268  &   82.28    &    1231.2   & 1211.5 \\ \hline
     &M3Y+Rep  & 	23.10	   & 7.365 &6.667  &0.328  &   65.01    &    1326.7    & 223.8 \\ \hline
\end{tabular}
\end{table} 

\begin{figure}[bht]\vspace{-2.0cm}
\hspace*{1.0cm}
\includegraphics[angle=0,scale=0.70]{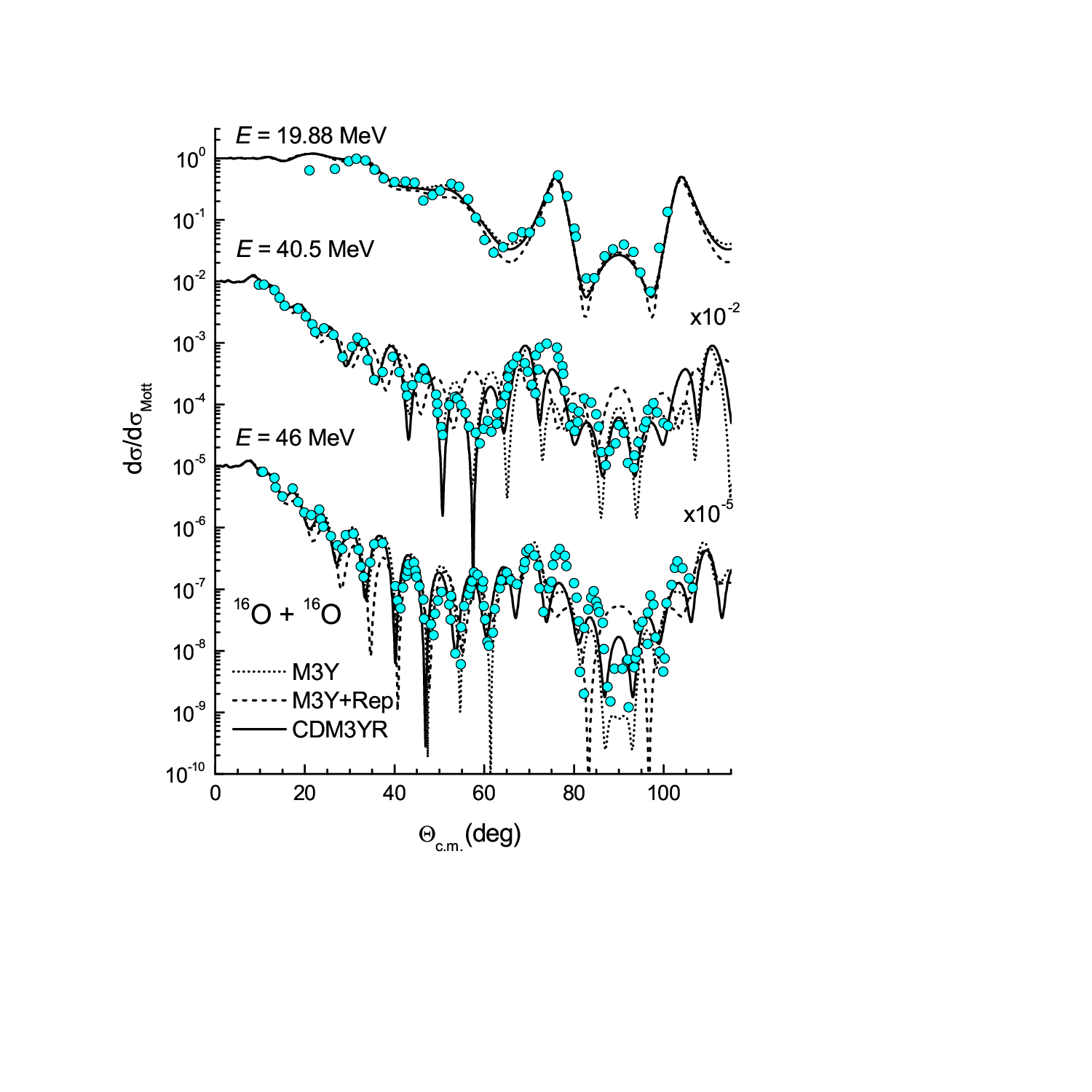}\vspace{-4cm}
\caption{OM description of  elastic \oo scattering at $E = 19.88, 40.5$, and 46 MeV given by three 
versions of the real folded OP. The WS parameters (\ref{eq18})-(\ref{eq19}) of the imaginary OP 
were adjusted in each case by the best OM fit to the measured elastic data \cite{Mat69,Nic99}.} \label{fig5}
\end{figure}
\begin{table}[bht]
\caption{\label{tOP2} WS parameters (\ref{eq18})-(\ref{eq19}) of the imaginary OP  obtained with 
three versions of the folded potential $V_{\rm N}$ from the best OM fit to elastic \oo scattering data
at $E=19.88, 40.5$, and 46 MeV. $J_R$ and $J_W$ are the volume integrals (per interacting 
nucleon pair) of the real and imaginary OP, respectively. $\sigma_R$ is the total reaction cross section 
given by the OM calculation. $\chi^2$ values are per datum, and obtained with uniform 
10\% errors.}\vspace{0.5cm}
\begin{tabular}{|c|c|c|c|c|c|c|c|c|c|c|c|c|} \hline
$E $ &Potential& $-J_R$  &$W_V$ & $R_V$ & $a_V$ & $W_D$ & $R_D$ & $a_D$  & $-J_W$ &$\sigma_R$  & $\chi^2$ \\
(MeV) &    & (MeV~fm$^3$)   & (MeV) & (fm)  & (fm)  & (MeV) & (fm)  & (fm)  &  (MeV~fm$^3$) & (mb)  & \\ \hline
19.88&CDM3YR    &  389.8      &10.460 &5.728 &0.292 &2.181&7.242&0.234&   38.26     &  992.8 & 14.0 \\ \hline
     &M3Y       &  484.4      &11.427 &5.637 &0.241 &1.988&7.253&0.256&   39.38     &  992.1 & 15.8 \\ \hline
     &M3Y+Rep   &  -73.27      &12.629 &5.386 &0.381 &1.927&7.526&0.322&   40.85     &  1096.3  & 13.5 \\ \hline
40.5 &CDM3YR    &  389.8      &13.469 &5.567 &0.203 &2.038&7.397&0.403&   47.44     &  1574.0  & 66.1 \\ \hline
     &M3Y       &  484.4      &15.462 &5.637 &0.333 &1.808&7.573&0.313&   53.29     &  1528.7 & 196.4 \\ \hline
     &M3Y+Rep   &  -73.27      &14.528 &5.496 &0.311 &2.352&7.483&0.321&   49.07     &  1508.3  & 765.0 \\ \hline
46.0 &CDM3YR    & 389.8       &17.463 &5.577 &0.313 &1.938&7.281&0.433&    59.94    &  1595.7 & 285.5 \\ \hline
     &M3Y       & 484.4       &16.462 &5.717 &0.332 &2.138&7.050&0.353&    59.45    &  1489.7 & 159.7 \\ \hline
     &M3Y+Rep   & -73.27       &16.130 &5.561 &0.510 &1.175&7.883&0.345&    54.15    &  1577.8  & 2347.1 \\ \hline
\end{tabular}
\end{table}

In Fig.~\ref{fig4}, the OM results obtained for elastic \cc angular distributions using three 
versions of the  folded nuclear potential $V_{\rm N}$ are compared with elastic data measured 
at energies near the Coulomb barrier \cite{Rei73}. Usually, a renormalization $N_r$ of the real folded 
OP is introduced and adjusted by the best OM fit to elastic scattering data \cite{Kho00,Kho07r}, and
the folding model can be well validated when the best-fit $N_r$ is close to unity. In the present work, 
we kept $N_r=1$ in all the OM and BPM calculations to explore in details the difference between 
three versions of the real folded OP.  In each case, the WS parameters of the imaginary  OP were 
adjusted by the OM fit to elastic \cc data, starting from parameters of the global OP for \cc system 
at higher energies \cite{McV92}, and the best-fit WS parameters are listed in Table~\ref{tOP1}.  

One can see in Fig.~\ref{fig4} that the three nuclear potentials based on CDM3YR, M3Y, and 
M3Y+Rep interactions provide more or less reasonable OM descriptions of  elastic \cc scattering 
at $E = 17.5$  MeV, because at this low energy elastic angular distribution is determined mainly 
by the Coulomb potential, and is sensitive to the nuclear potential only at the surface where three folded 
potentials are quite close in shape (see Fig.~\ref{fig3}). At higher energies, the broad oscillation 
of elastic cross section around $\theta_{\rm c.m.}\approx 90^\circ$ becomes more sensitive 
to the real OP at sub-surface distances, and the difference of  three potentials begins to show up in 
the calculated elastic cross section. Although the $\chi^2$ values (see Table~\ref{tOP1}) are of about 
the same order, only the CDM3YR potential consistently reproduces the magnitude and oscillation 
pattern of the 22.5 and 32.5 MeV data at angles $\theta_{\rm c.m.} \gtrsim 60^\circ$. 

The OM results obtained for elastic \oo scattering using three folded potentials $V_{\rm N}$ are 
compared in Fig.~\ref{fig5} with elastic data measured at energies $E = 19.88, 40.5$, and 46 MeV,
over a wide angular range, up to $\theta_{\rm c.m.} \simeq 100^\circ$ \cite{Mat69,Nic99}. At each 
energy, the WS parameters (\ref{eq18})-(\ref{eq19}) of the imaginary OP (see Table~\ref{tOP2}) were 
adjusted by the best OM fit to the data, starting from the \oo global OP adopted at higher energies 
\cite{Nic99}. At low energy of 19.88 MeV, the measured elastic data are sensitive mainly to Coulomb 
potential and far tail of the nuclear potential, and all three real folded potentials give about 
the same OM description of the measured data. Like the \cc case, the difference in three folded potentials 
shows up at higher energies of 40.5 and 46 MeV, especially, at angles $\theta_{\rm c.m.} \gtrsim 70^\circ$, 
where both the deep M3Y and shallow M3Y+Rep potentials fail to describe the magnitude and oscillation 
pattern of the measured elastic cross section. The $\chi^2$ values given in Table~\ref{tOP2} represent 
an average deviation of the calculated elastic cross section from the measured data over the whole
angular range, and those obtained with the M3Y+Rep potential turned out to be an order of magnitude 
larger than those obtained with the CDM3YR potential at  $E=40.5$ and 46 MeV. This is a clear 
evidence on the inadequacy of the shallow (repulsive) M3Y+Rep potential.    
  
We note further that symmetric \cc and \oo systems were proven to be strongly refractive, with the dominant 
farside scattering at medium and large angles \cite{Kho07r,Phu24}. Given the boson symmetry of these 
systems, the symmetric Mott interference pattern at scattering angles around $\theta_{\rm c.m.}=90^\circ$ 
was shown to be originated from an interference of the direct- and exchange components of the farside 
scattering amplitude and is, therefore, strongly sensitive to the real OP at sub-surface distances (see more 
details in Ref.~\cite{Phu24}). Built upon the realistic density dependent M3Y-Reid interaction 
(\ref{eq3})-(\ref{eq4}), the real folded CDM3YR potential gives consistently a good OM description 
of the data measured in the large-angle region around $90^\circ$ for both the \cc and \oo systems.

The volume integrals (per interacting nucleon pair) $J_R$ and $J_W$ of the real and imaginary OP 
are given explicitly in Tables~\ref{tOP1} and \ref{tOP2} to show the gross \emph{attractive} and 
\emph{absorptive} strengths of the real and imaginary OP, respectively. It can be seen that the M3Y 
potential, based on the density independent M3Y-Reid interaction (\ref{eq3}), is too attractive 
with $J_R$ of -470 to -480 MeV~fm$^3$. With a realistic density dependence (2) added to the original 
M3Y-Reid interaction (\ref{eq3}) to reproduce the saturation of symmetric NM as shown in 
Fig.~\ref{fig1}, the folded CDM3YR potential becomes less attractive with $J_R$ of  -380 to 
-390 MeV~fm$^3$, in a nice agreement with the empirical range of  $J_R$ established 
for \cc and \oo systems (see, e.g.,  Fig. 6.7 in Ref.~\cite{Bra97}). The M3Y+Rep potential (roughly 
the real folded M3Y potential added by a repulsive core) is very weakly attractive in the \cc case, 
with $J_R$ around -23 MeV~fm$^3$. For the \oo system, the M3Y+Rep potential becomes even 
grossly repulsive, with a positive $J_R$ around 73 MeV~fm$^3$ (see Table~\ref{tOP2}). Given similar 
absorptive strengths by the (best-fit) imaginary WS potentials, we conclude that the difference in the OM 
description of elastic \cc and \oo scattering shown in Figs.~\ref{fig4} and \ref{fig5} is mainly due to 
the difference between the three real folded nuclear potentials $V_{\rm N}$.   

\section{Astrophysical $S$ factors of \cc and \oo fusions}
\label{sec3} 
The fusion reaction of symmetric \cc and \oo systems at sub-barrier energies plays an important role in 
nuclear astrophysics studies of the origin of heavy elements, known as the carbon and oxygen burning chains 
in massive stars. Within the BPM \cite{Bal98}, the nuclear fusion at such low energies occurs via quantum 
tunneling of  the incoming wave through the $\ell$-dependent barrier determined by a counterbalance of the 
\emph{attractive} nuclear potential and \emph{repulsive} Coulomb- and centrifugal potentials
\begin{equation}
 V_\ell (R)= V_{\rm N}(R)+V_{\rm C}(R)+\frac{\hbar^2\ell(\ell+1)}{2\mu R^2}. \label{eq20}
\end{equation}
The fusion cross section is then estimated by summing the partial waves series of the penetration coefficient 
$T_\ell$ as 
\begin{equation}
 \sigma_{\rm f} = \frac{\pi}{k^2} \sum_\ell (2\ell+1)[1+ (-1)^\ell] T_\ell, \label{eq21}
\end{equation}
where $k$ is the relative-motion momentum and $[1+ (-1)^\ell]$ factor is implied by the boson symmetry 
of  identical \cc and \oo systems. 

For all partial waves $\ell$ with the barrier height $V_{\rm B \ell}$ lying above the c.m. energy  
($V_{\rm B \ell} \geq E$), $T_\ell$ for sub-barrier fusion is given by the WKB approximation \cite{Gas05} as 
\begin{equation}
 T_\ell =\frac{T_\ell^{\rm WKB}}{1 + T_\ell^{\rm WKB}}\qquad {\rm with} \qquad
 T_\ell^{\rm WKB}=\exp\left\{-\frac{2}{\hbar}\int^{R_2}_{R_1}
 \sqrt{2\mu[V_\ell(R)-E]}{\rm d}R\right\}, \label{eq22}
\end{equation}
where $R_1$ and $R_2$ are the inner and outer turning points with $V_\ell(R_1) = V_\ell(R_2) = E$. 
 
When the c.m. energy is higher than the barrier height ($V_{ \rm B \ell} < E$), the well-known 
Hill-Wheeler formula \cite{Hil53} is used to determined $T_\ell$ 
\begin{equation}
T_\ell= \left\{1 + \exp\left[\frac{2\pi (V_{\rm{B}\ell}-E)}{\hbar \omega_\ell} \right] \right\}^{-1}, 
\label{eq23}
\end{equation}
where $\hbar\omega_\ell$ is the curvature of the total potential (\ref{eq20})  at the barrier-height radius 
$R_{\rm{B}\ell}$
\begin{equation}
 \hbar\omega_\ell=\bigg |\frac{\hbar^2}{\mu}\frac{{\rm d}^2 V_\ell(R)}{{\rm d}R^2}
 \bigg |_{R=R_{\rm{B}\ell}}^{1/2} \ \ {\rm with} \ \ 
 V_{\rm{B}\ell}=V_\ell(R_{\rm{B}\ell}). \label{eq24}
\end{equation}
\begin{figure}[bht]\vspace{0cm}
\hspace*{0cm}
\includegraphics[angle=0,scale=0.50]{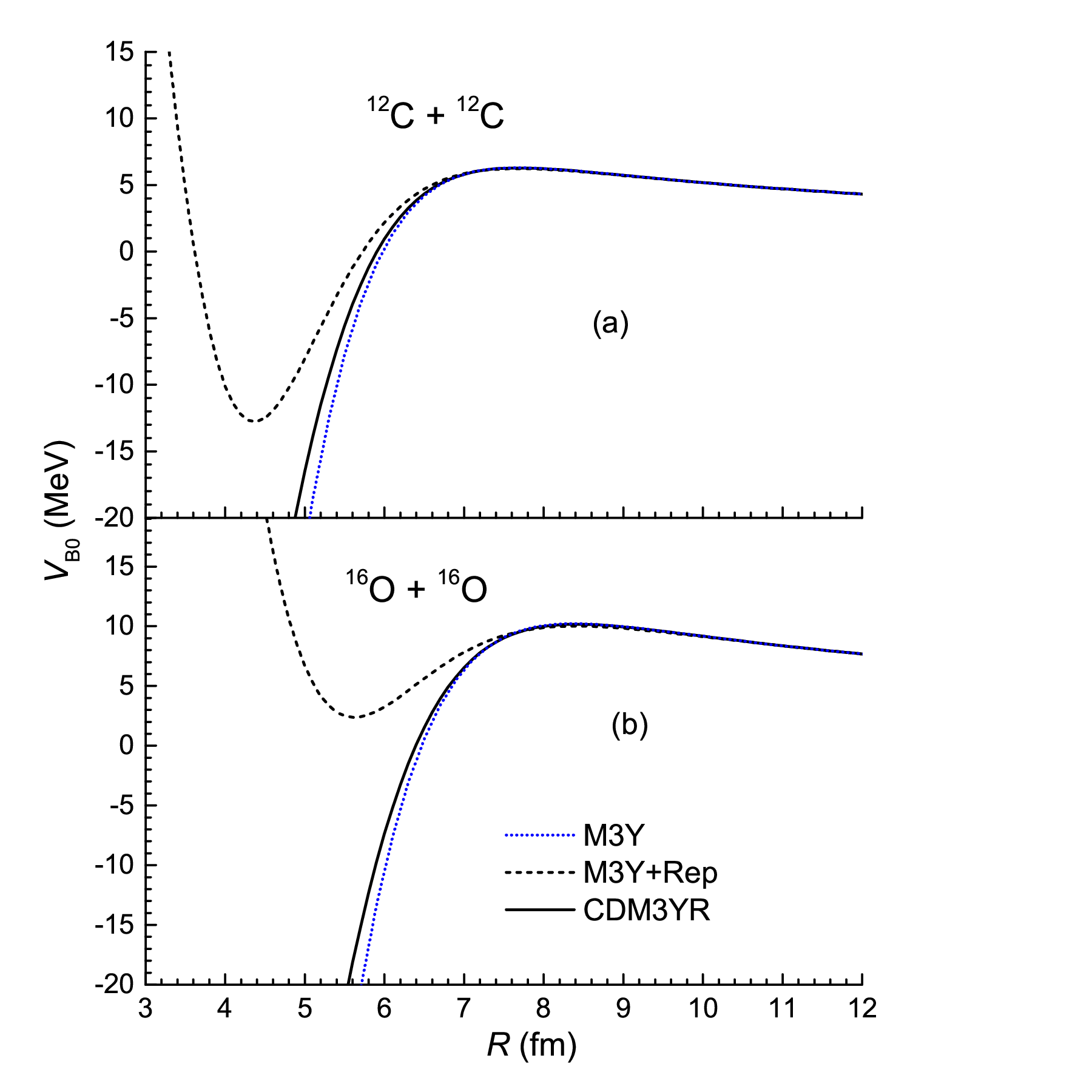}\vspace{0cm}
\caption{Radial shape of the potential barrier \eqref{eq20} obtained with three nuclear potentials 
for the $\ell=0$ partial wave of  \cc (a) and  \oo systems (b) . The solid, dashed, and dotted curves 
are given by the CDM3YR, M3Y+Rep, and M3Y folded potentials, respectively.} \label{fig6}
\end{figure}

\begin{figure}[bht]\vspace{0cm}
\hspace*{0cm}
\includegraphics[angle=0,scale=0.50]{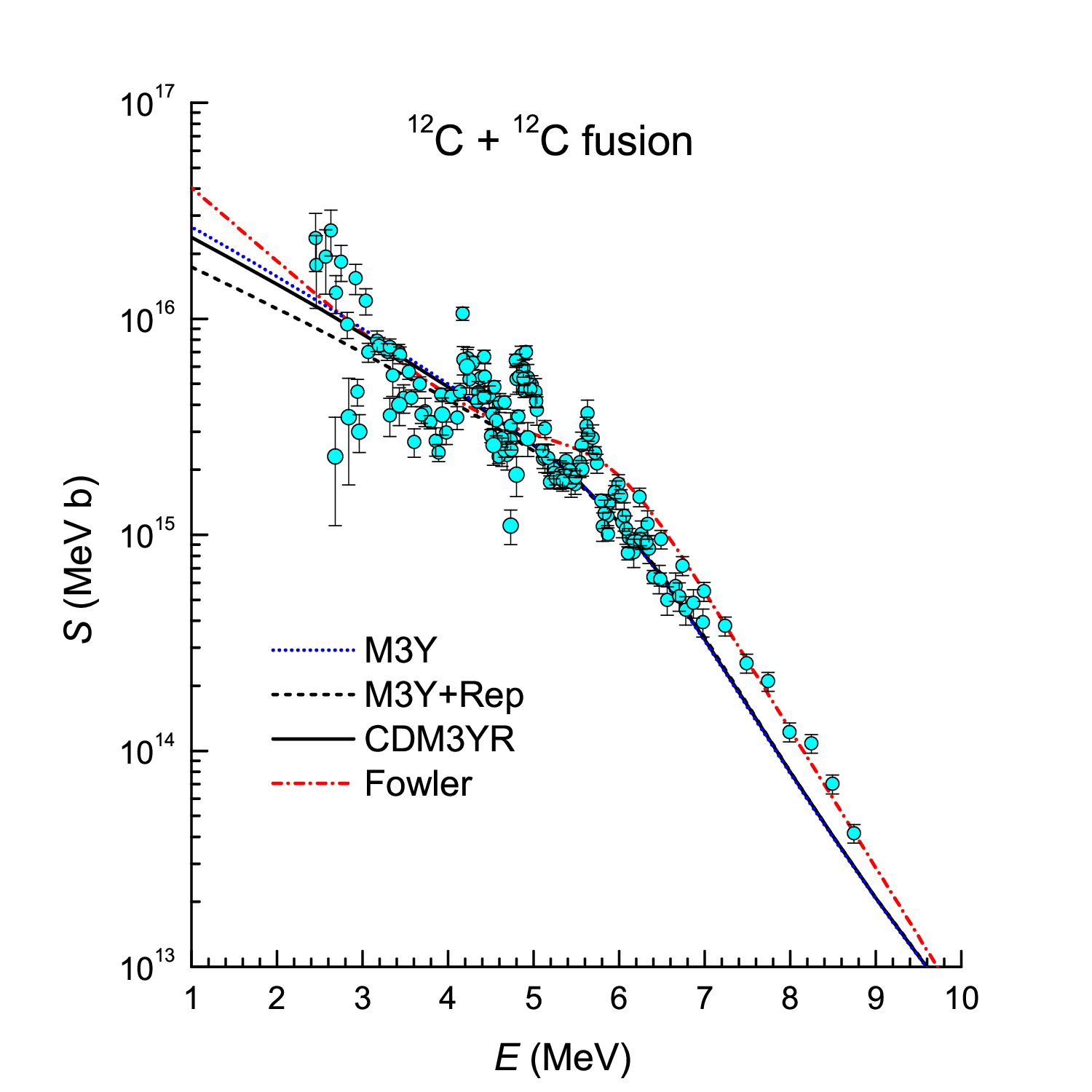}\vspace{0cm}
\caption{Astrophysical $S$ factor \eqref{eq25} for \cc system. The solid, dashed, and dotted curves are given 
by the CDM3YR, M3Y+Rep, and M3Y folded potentials, respectively. The dashed-dotted curve is given by 
Fowler's empirical parametrization \cite{Fow75}. Data are taken from  
Refs.~\cite{Pat69,Maz73,Hig77,Das82,Agu06,Bar06,Spi07,Jia18}.} \label{fig7}
\end{figure}
\begin{figure}[bht]\vspace{0cm}
\hspace*{2cm}
\includegraphics[angle=0,scale=0.50]{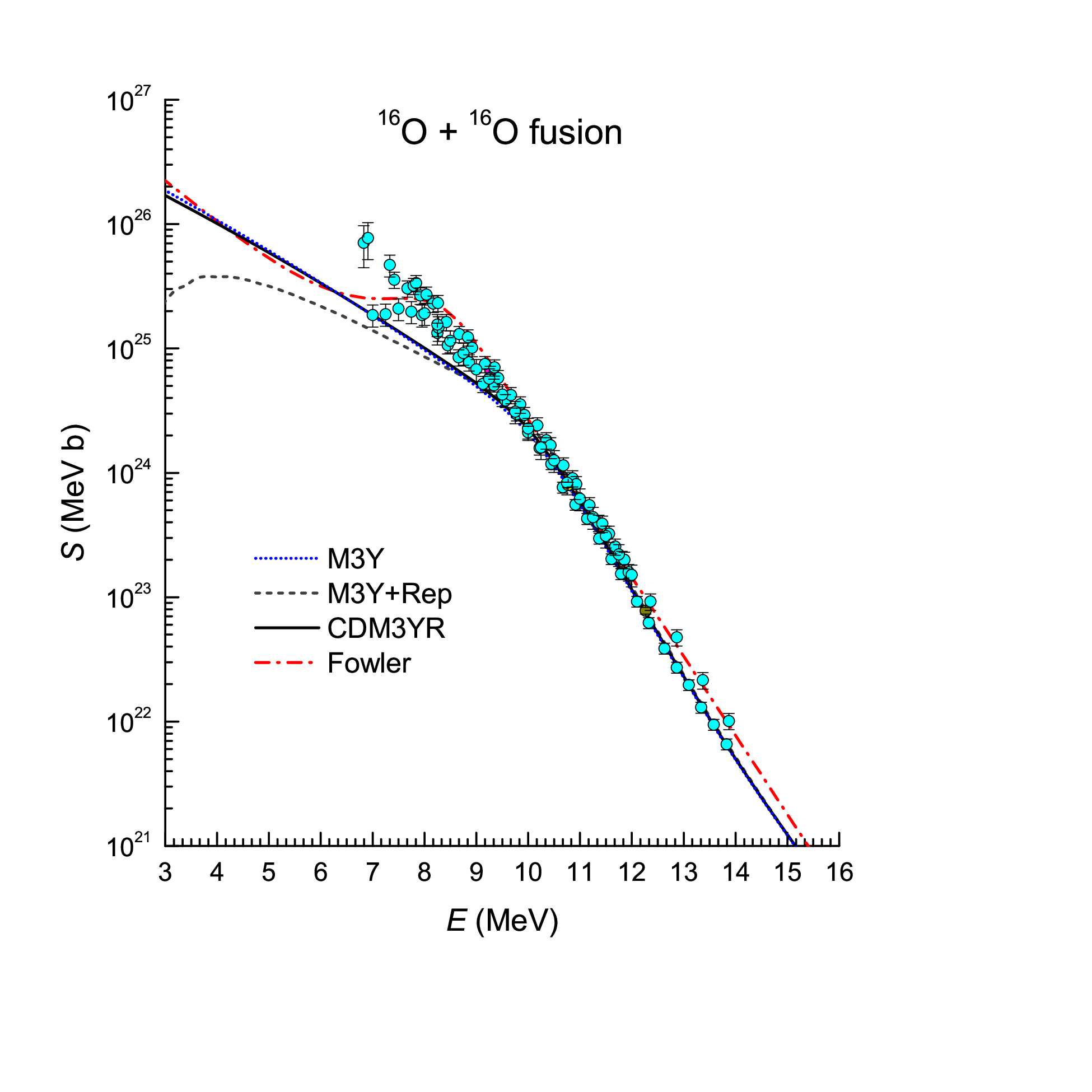}\vspace{-2.0cm}
\caption{The same as Fig.~\ref{fig7} but for \oo system. Data are taken from 
Refs.~\cite{Spi74,Hul80,Wu84,Kur87,Dua15}.} \label{fig8}
\end{figure}

Because the tunneling probability decreases exponentially with the decreasing energy, the fusion cross 
section drops too rapidly at energies below the Coulomb barrier, and this poses a big challenge for 
empirical studies to extrapolate the fusion cross section from higher energies. At astrophysical energies, 
it is convenient to introduce the astrophysical $S$ factor that varies less rapidly with the decreasing energy
\begin{equation}
 S = E e^{2\pi \eta} \sigma_{\rm f}, \label{eq25}
\end{equation}
where $\eta =Z_1 Z_2 e^2/(\hbar v)$ is the Sommerfeld parameter, and $v$ is the relative velocity 
of the dinuclear system. In particular, $2 \pi \eta\approx 87.2/\sqrt{E~ {\rm (in\ MeV)}}$ {and} 
$179.0/\sqrt{E~{ \rm (in\ MeV)}}$ for \cc and \oo systems, respectively.

Because the repulsive Coulomb and centrifugal potentials in Eq.~\eqref{eq20} are well determined, 
a realistic nuclear potential is the vital input for the BPM calculation of the astrophysical $S$ factor. 
In the present fusion study, the three nuclear folded potentials were used as the nuclear term $V_{\rm N}$ 
in Eq.~(\ref{eq20}). It is obvious from Eqs.~\eqref{eq23} and \eqref{eq24} that the fusion cross section 
$\sigma_{\rm f}$ should be sensitive mainly to the strength and slope of the nuclear potential around 
the Coulomb barrier. For illustration, we have plotted in Fig.~\ref{fig6} the radial shape of  the 
potential barrier $V_{\rm B0}$ obtained with three folded potentials for $\ell=0$ partial wave of  \cc and 
\oo systems. One can see that all three nuclear folded potentials provide about the same barrier heights and 
radius, $V_{\rm B0}\approx 6.2$ MeV and $R_{\rm B0}\approx 7.8$ fm for \cc system; $V_{\rm B0}\approx 
10.2$ MeV and $R_{\rm B0}\approx 8.4$ fm for \oo system. These values agree reasonably with the empirical 
estimates \cite{Azi15,Agu06}. The three folded potentials also give similar barrier shapes at distances beyond 
the barrier-height radius. However, at sub-barrier distances, the repulsive M3Y+Rep potential gives a 
significantly different radial shape on the inner side of potential barrier compared to that obtained with 
the attractive M3Y and CDM3YR potentials. Such a difference of the sub-barrier radial shape shows 
up in the BPM prediction of astrophysical $S$ factor at lowest energies, as illustrated in Fig.~\ref{fig7} 
and \ref{fig8}. 

From the results shown in Fig.~\ref{fig7} for \cc system one finds that  two attractive M3Y and CDM3YR 
potentials deliver about the same astrophysical $S$ factors over energies down to Gamow window of about 
1.5 MeV, while the $S$ factor given by the M3Y+Rep potential is substantially lower at these energies, 
especially, when compared with that given by Fowler's empirical parametrization \cite{Fow75}. Such a 
different behaviors of the $S$ factor is caused by the repulsive core of weakly attractive M3Y+Rep 
potential (see Table~\ref{tOP1}). Nevertheless, the repulsive core of the M3Y+Rep potential of \cc system
is not strong enough to bend down the curve of the $S$ factor at the lowest energies as might be implied 
by the hindrance of \cc fusion, a still debated issue \cite{God19,Uza25}. 

Because of larger Coulomb potential combined with a grossly repulsive M3Y+Rep potential 
(see Table~\ref{tOP2}), the difference in the potential barrier given by the M3Y, CDM3YR and 
M3Y+Rep nuclear potentials for \oo system is more substantial at distances from the inner turning 
point $R_1$ to the barrier-height radius $R_{\rm B0}$. This leads to a drastic difference in the calculated 
astrophysical $S$ factor of \oo fusion shown in Fig.~\ref{fig8}. At sub-barrier energies, the $S$ 
factor given by the M3Y+Rep potential saturates at a broad maximum near Gamow window of about 
4 MeV, and drops further at lower energies. Such a behavior of  the $S$ factor given by the M3Y+Rep 
potential for \oo fusion at sub-barrier energies was also found and linked to a possible hindrance 
of \oo fusion at low energies \cite{Esb08}. In general, the fusion hindrance remains an open question, 
but within the present potential model study the hindrance of  \oo fusion is simply due to a strong 
repulsive core of the M3Y+Rep potential used in the one-channel BPM calculation. Likewise, 
the recent OM calculation using a strongly repulsive real OP also obtained similar hindrance 
of \oo fusion \cite{Islam25}. Because the repulsive core of the M3Y+Rep potential is added in a 
phenomenological manner \eqref{eq16} which is still questionable as discussed in Sec.~\ref{sec2}, 
it remains highly uncertain whether the hindrance of \oo fusion can be predicted in such a potential model 
calculation. For this purpose, a multi-channel microscopic study of the main particle-emission channels 
of \oo fusion like $^{16}$O($^{16}$O,$\alpha)^{28}$Si... is highly desirable, as carried out recently 
by Descouvemont for \cc system \cite{Des26}.  
               
\section{Summary}
A realistic CDM3YR density dependence has been added  to the original (density independent) 
M3Y-Reid interaction \cite{Ber77} to properly reproduce the empirical saturation point of cold NM
in the nonrelativistic HF calculation, with the nuclear incompressibility $K\approx 223$ MeV. 
The mean-field HF results obtained for energy of NM agree well with those given by the \emph{ab-initio} 
APR calculation \cite{Akm98} over a wide range of NM densities, up to three times the saturation 
density $\rho_0$. 

The reliability of the newly parametrized density dependent CDM3YR interaction has been 
probed in the DFM calculation of the real OP for the OM study of elastic \cc and \oo 
scattering at low energies, as well as the BPM calculation of the astrophysical $S$ factor 
of \cc and \oo fusion. The impact of nuclear medium to the strength and shape of the real OP 
is well taken into account in the DFM calculation using the density dependent CDM3YR 
interaction, and an appropriate adiabatic density approximation for the \AA overlap density.

For comparison, the same OM calculations were also done with the real folded OP given by 
the original (density independent) M3Y-Reid interaction \cite{Ber77}, and that added by a repulsive 
core proposed by Esbensen {\it et al.} \cite{Mis06,Mis07,Esb08,Esb11}. Without further renormalizing 
the potential strength, only the CDM3YR potential delivers consistently a good OM description 
of elastic \cc and \oo scattering at low energies, especially, the data points measured around 
$\theta_{\rm c.m.}=90^\circ$ that were proven to be sensitive to the interior of the real OP. 
The fact that both the M3Y and M3Y+Rep folded potentials fail to describe the magnitude 
and oscillation pattern of elastic data measured at medium and large angles 
indicates the unrealistic shape of these two potentials at sub-surface distances. 
 
Three versions of the (folded) nuclear potential were further used in the BPM study of \cc and \oo fusion 
down to astrophysical energies that are not directly accessible by experiment. Within the one-channel 
BPM, the quantum tunneling is determined by the nuclear potential mainly around the Coulomb barrier, 
and existing fusion data are usually not that sensitive to the shape of the real OP, compared to elastic 
scattering data measured for the same system. The difference in the potential shape at sub-barrier 
distances shows up, however, in the calculated astrophysical $S$ factor at low energies down to 
Gamow window. Here, the difference between the repulsive M3Y+Rep potential and attractive 
M3Y and CDM3YR potentials on the inner side of the Coulomb barrier leads to a sizable difference 
of the calculated astrophysical $S$ factor. The strong repulsive core of the M3Y+Rep potential 
of \oo system even gives rise to the fusion hindrance, which was not found in the BPM result 
given by the M3Y+Rep potential of \cc system. Given the potential ambiguity discussed 
in the present study, it is highly uncertain if the fusion hindrance of \cc or \oo system 
can be simply validated in the one-channel BPM calculation.     

\section*{Acknowledgments}
The present research has been supported, in part, by the National Foundation for Science and 
Technology Development (NAFOSTED Project No. 103.04-2025.06). 

\section*{Data availability}
The data that support the findings of this article are openly available \cite{Rei73,Mat69,Nic99,
Pat69,Maz73,Hig77,Das82,Agu06,Bar06,Spi07,Jia18,Spi74,Hul80,Wu84,Kur87,Dua15}.

\bibliographystyle{apsrev4-2}
\bibliography{references}
\end{document}